\documentclass{llncs}

\usepackage{macros}

\title{Point-and-write --- Documenting Formal Mathematics by Reference\thanks{The first and third author were funded by the NWO project ``MathWiki''. The second author was supported by DFG Project I1-[OntoSpace] of SFB/TR 8 “Spatial Cognition” and EPSRC grant EP/J007498/1''.  The final publication is available at \texttt{http://www.springerlink.com}.}}
\author{Carst Tankink\inst{1}
   \and Christoph Lange\inst{2,3,4} 
   \and Josef Urban\inst{1}}
\institute{Institute for Computing and Information Science, Radboud Universiteit, Nijmegen, The Netherlands \email{carst@cs.ru.nl}, \email{josef.urban@gmail.com}
     \and FB 3, Universität Bremen, Germany \email{ch.lange@jacobs-university.de}
     \and Computer Science, Jacobs University Bremen, Germany
     \and Computer Science, University of Birmingham, UK}

\begin{document}
\maketitle

\begin{abstract}
  This paper describes the design and implementation of mechanisms for light-weight
  inclusion of formal mathematics in informal mathematical writings,
  particularly in a Web-based setting. This is conceptually done in
  three stages: (i) by choosing a suitable representation layer (based
  on RDF) for encoding the information about available resources of
  formal mathematics, (ii) by exporting this information from formal
  libraries, and (iii) by providing syntax and implementation for
  including formal mathematics in informal writings.

  We describe the use case of an author referring to formal text from an informal narrative, and discuss design choices entailed by this use case. Furthermore, we describe an implementation of the use case within the Agora prototype: a Wiki for collaborating on formalized mathematics.
\end{abstract}

\section{Introduction}
\label{sec:introduction}

Formal, computer-verified, mathematics has been informally discussed and written about for some fifty
years: on dedicated mailing
lists~\cite{CoqClub,MizarForum,IsabelleUsers}, in
conference and journal 
articles, online manuals, tutorials and courses, and in community
Wikis~\cite{Cocorico,MizarTWiki} and 
blogs~\cite{Univalence}.

In such informal writings, it is common to
include and mix formal definitions, theorems, proofs and their
outlines, and sometimes whole sections of
formal articles. Such formal ``islands'' in a text do
not have to follow any particular logical order, and can mix content from
different articles, libraries, and
even content based on different proof assistants. In this respect, the
collection of such formal fragments in a particular text is often
\emph{informal}, because the fragments do not have to share and form a
unifiable, linear, and complete \emph{formal context}.

In a Web setting, such pieces of formal code can however be equipped with 
semantic and presentation functions that make formal
mathematics attractive and unique. Such functions range from
``passive'' markup, like (hyper)linking symbols to their precise definitions in
HTML-ized formal libraries, detailed and layered
explanations of implicit parts of reasoning (goals, types, subproofs,
etc.), to more ``active'', like direct editing, re-verification, and
HTML-ization of the underlying formal fragment in its proper context, and using the formal code
for querying semantic search engines and automated reasoning
tools~\cite{UrbanS10}.

In this paper, we describe, and support, a use case of an author writing such an informal text: she gives references (\demph{points}) to (fragments of) formalization on the Web, and then describes (\demph{writes} about) them in a natural language narrative, documenting the formal islands. This use case is described in Section~\ref{sec:using}.

We support this use case in a light-weight manner, based on HTML presentations of formal mathematics. The author can write pointers to formal objects in a special syntax (described in Section~\ref{sec:syntax}), which get resolved to the objects when rendering the narrative. 
To follow the pointers, our tools equip HTML pages for formalizations with annotations describing \emph{what} a particular HTML fragment represents (Section~\ref{sec:annotating}).
The annotations are drawn from suitable RDF vocabularies,\footnote{The RDF data model (Resource Description Framework) essentially allows for identifying any thing (``resource'') of interest by a URI, giving it a type, attaching data to it, and representing its relations to other resources~\cite{w3c:rdf-concepts}.} described in Section~\ref{sec:types}.

We show an implementation of the mechanisms in the Agora prototype,\footnote{\url{http://mws.cs.ru.nl/agora/}} described in Section~\ref{sec:system}.
The actual \emph{rendering} of the final page with its inclusions give rise to several issues that we do not consider here: we discuss some issues and how we handle them in our implementation, but these decisions were not made systematically: we focus here on the author's use case of writing the references, and provide the prototype as a proof of existence.

This paper does not contain a dedicated related work section: to our knowledge, there is no system that provides similar functionality: there are alternatives for including formal text, as well as alternative syntaxes, which we will compare with our approach in the relevant sections. Additionally, the MoWGLI project developed some techniques for rendering formal proofs with informal narratives, which are compared to in Section~\ref{sec:system}.
\section{Describing \emph{and} Including Formal Text}
\label{sec:using}

The techniques described in this paper are mainly driven by a single use case, that of an author writing a description (a ``narrative'') of a development in formal, computer-verified, mathematics. In this work, we assume that the author writes this narrative for publication in a Wiki, although the use case could also be applied for more traditional authoring, in a language like \LaTeX.

\subsection{Use Case}
While writing a natural language narrative, the author will eventually want to include snippets of formal code: for example to illustrate a particular implementation technique or to compare a formalization approach with a different one, possibly in a different formal language. An advanced example is Section 5.3 of~\cite{BancerekR02} rendered in Agora.\footnote{\url{http://mws.cs.ru.nl/agora/cicm_sandbox/CCL/}}

Because we allow for including formal text from the Web, there is a wrinkle we have to iron out when supporting this use case, but before we get to that, we describe typical steps an author can carry out while executing the use case:

\begin{description}
  \item[Formalization] An author works on a formalization effort in some system and puts (parts of) this formalization on the Web, preferably on the Wiki. We will refer to the results of these efforts as \demph{source texts}.
  \item[Natural language description] Sometime before, during or after the formalization, the author gives a natural language account of the effort on the Wiki. As mentioned, we assume she writes this in a markup language suitable for Wikis, extended with facilities for writing mathematical formulae.\footnote{The specifics of suitable languages for writing in Wikis for formal mathematics are not a subject of this paper; we refer to~\cite{MathWiki11} for an overview.} We refer to the resulting description as a \demph{narrative}.
  \item[Including formalizations into the narrative] In the natural language description, the author includes some of the formal artifacts of her effort, and possibly some of the formalizations by other authors. These inclusions need to look attractive (by being marked up) and should not be changed from the source: the source represents a verifiable piece of mathematics, and a reader should be able to ascertain himself that nothing was lost in transition. 
\end{description}

These steps are not necessarily carried out in order, and can be carried out by different authors or iteratively. In particular, the formal text included in the narrative does not have to originate from the author or her collaborators, but could be from a development that serves as competition or inspiration.

\begin{figure}
\includegraphics[width=.8\textwidth]{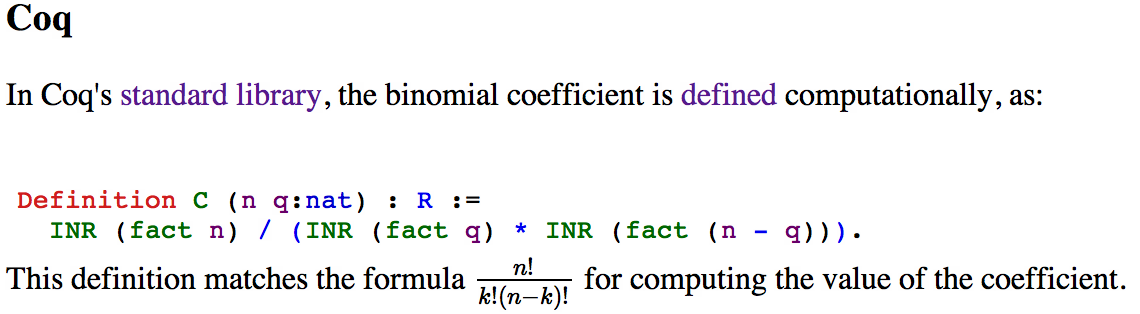}
\caption{Example of informal narrative with formal snippets}
\label{fig:result}
\end{figure}

The end results of the workflow are pages like the one shown (in part) in Figure~\ref{fig:result}: it includes a narrative written in natural language (including hyperlinks and markup of formulae) and displays formal definitions marked up as code.

Because the source texts are stored on the Web, we consider their content to be \emph{fluid}: subject to change at any particular moment, but not under control of the author. This implies that the mechanism for including formal text should be robust against as much change as possible.

To determine how we can support this workflow, we first survey the existing methods that are suitable for including formal mathematics. 

\subsection{Alternatives for Inclusion}
Typical options for including formal mathematics---or any other type of code\mbox{---,} when working with a document authoring tool like \LaTeX\ include: 
\begin{enumerate}
  \item \textbf{Referral:} place the code on some Web page and refer readers to that page from the document (by giving the URL),
  \item \textbf{Inclusion:} \label{it:listings} include and format the source code files as listings, e.g.\ using the \LaTeX{} package \tool{listings}~\cite{ctan/macros/latex/contrib/listings:oldfashioned}, 
  \item \textbf{Literate proving:} the more extreme variant of (\ref{it:listings}): write the article in a literate style, and extract both formal code and marked up text from it,
  \item \textbf{Copy-paste:} manually copy-paste the code into the document.
\end{enumerate}

All of these options have their own problems for our use case.
\begin{enumerate}
  \item Referral collides with the desire for \emph{juxtaposability}~\cite{BlackwellGreen-1998}: a reader should not have to switch between pages to look at the referred code and the text that refers to it. Instead, he should be able to read the island within the context of the narrative.
  
  \item We certainly want the author to be able to include code, but most of the tools only allow her to refer to the code by \emph{location}, instead of a more semantic means: she can give a range of lines (or character offsets) in a file, but cannot write ``include the Fundamental Theorem of Algebra, and its proof''.
  \item Literate proving~\cite{CG:LiterateProving05} is a way to tackle code inclusion, but it does not solve the use case: it requires the author to shift her methods from writing code and article separately to writing both aspects interleaved. 
  
  It also does not allow an author to include existing external code for citation (without copy-pasting) and does not allow her to write a document including only snippets of formal code. These cases can arise where a lot of setup and auxiliary lemmas are necessary for formalizing a theorem, but only the theorem itself is the main focus of a paper. Typical literate programming setups provide mechanisms for hiding code fragments, but we prefer to take an inclusive instead of an exclusive view on the authoring process: the former seems to be more in line with actual practices in the interactive theorem proving community (see, for example, the proceedings of the Interactive Theorem Proving conference~\cite{VanEekelen+-2011}).
  \item Copy-pasting code has the traditional problem of maintaining consistency: if the source file is changed, the citation should change as well. On the positive side, it does not require much effort to implement, apart from adding facilities for marking up code, which can be reused for in-line (new) code. To make the implementation threshold even lower, the \tool{listings} \LaTeX\ package previously mentioned also supports marking up copy-pasted code. 
\end{enumerate}

The shortcomings of these methods mean we need to design a system providing the following facilities:

\begin{requirement}
A \emph{syntax} for writing, in a natural language document, references to parts of a formal text, possibly outside of the referring text, and a mechanism for 
including the referred objects verbatim in a rendered version of the natural language text.
\end{requirement}

\begin{requirement}
A method for 
annotating parts of formal texts, so they can be referenced by narratives.
\end{requirement}

The rest of this paper gives our approach to these two problems, demonstrating how they interact, and gives a tour of our working implementation.\footnote{\url{http://mws.cs.ru.nl/agora/cicm_sandbox}}

\section{Syntax for Referring}
\label{sec:syntax}
We will first focus on \emph{how} the author can write references to formal content.  Below, we discuss considerations that guided the design of this syntax. The considerations are partially based on the goals for a common Wiki syntax~\cite{Sauer+-2007}. 

\subsection{Requirements on syntax}
\label{subsec:syntax-requirements}
\begin{description}
  \item[Simple] To encourage its use, the syntax should not be too elaborate. An example of a short enough syntax is the hyperlink syntax in most Wiki systems: only four characters surrounding the link, \verb|[[| and \verb|]]|. 
  \item[Collision free] The syntax should not easily be `mistyped': it should not be part of the syntax already used for markup, and not likely used in a natural language narrative.
  \item[Readable] It should be recognizable in the source of the narrative, to support authors in learning a new syntax and making the source readable.
  \item[Familiar] We do not intend to reinvent the wheel, but want to adapt existing syntax to suit our needs. This also keeps the syntax readable: when the base syntax is already known to an author, it should be clear to her how this syntax works in the context of referring to formal text. Because keeping things similar but not completely equal could cause confusion, it also requires us not to deviate the behavior too much from the original syntax.
\end{description}

In resolving these requirements on the syntax, we need to consider the context in which it will be used. In our proposed use case, the syntax will be used within a Wiki, so we prefer a syntax that fits with the markup families used for Wiki systems. It should be possible to extend a different markup language (for example, \LaTeX\ or literate comments for a formal system) with the reference syntax, but this requires reconsidering the decisions we make here.

Considering that we want to base the reference syntax on existing mechanisms (in line with the familiarity requirement), there are three basic options to use as a basis: import statements like used in \LaTeX\ (or, programming and formal languages), Wiki-style hyperlinks, and \tool{Isabelle/Isar}'s antiquotation syntax.

Each of these is considered in the rest of this section, and tested against the requirements stated before.

\paragraph{\LaTeX-style include statements.} The purpose of these statements in a \LaTeX\ document is to include the content of a file in another, before rendering the containing file. The command does not allow inclusion of file fragments, but could be modified to allow this. As a concrete proof of existence, the \tool{listings} package mentioned earlier has the option to include file fragments by giving a line offset, but not a pointer to an object. The statements should be recognizable by \LaTeX\ users or users of a formal language that uses inclusions, but the statements are rather long: if the author wants to use them more often, it might become tedious to write.

The MediaWiki engine has a similar syntax for including entire pages.\fnurl{http://www.mediawiki.org/wiki/Transclusion} 
To include fragments of pages, however, one either needs to factor out these fragments of the source text and include them both in the source text and the referring text, or mark fragments of the source page which will be included. Both do not give the author of a narrative fine-grained control over inclusion.

An extension to these inclusions\fnurl{http://www.mediawiki.org/wiki/Extension:Labeled_Section_Transclusion} allows inclusion of sections. This mechanism is a valid option for adaption, but if we would want to support informal inclusion at some point, it would be difficult to distinguish it, at the source level, from an inclusion of a formal fragment.
\paragraph{Wiki-style hyperlinks.} These cross-reference statements are not hard to learn and short, but also using them for inclusion can overload the author's understanding of the markup commands: if she already knows how to use hyperlinks, she needs to learn how to write and recognize links that include formal objects.

\paragraph{Antiquotations.} \tool{Isabelle/Isar}~\cite{Wenzel-2002} uses antiquotations to allow the author of \tool{Isar} proof documents to write natural-language, marked-up snippets in a formal document (the `quotation' from formal to informal), while including formal content in these snippets (the `antiquotation' of informal back to formal): these antiquotations are written {\st @\{type [options] syntax \}}, where {\st type} declares what kind of syntax the formal system can expect, the {\st syntax} specifying the formal content, and the {\st options} defining how the results should be rendered. The formal system interprets these snippets and reinserts the results into the marked-up text.

For example, the antiquotation {\verb#@{term [show_types] "% x y. x"}#} would ask \tool{Isabelle} to type-check the term {\st $\lambda x\ y . x$} (\% is \tool{Isabelle}'s ASCII shorthand for $\lambda$) in the context where it appears and reinserts the term annotated with its type: it inserts the output {\st $\lambda$(x::'a) y::'b.\ x}.

Another example is {\st @\{thm foo\}} which inserts the statement of the theorem labeled {\st foo} in the marked up text. The syntax also provides an option to insert the label {\st foo}, which makes sure that it points to a correct theorem.

\subsection{Resulting Syntax}
From the options listed above, the antiquotation mechanism is closest to what we want: it allows the inclusion of formal text within an informal environment, relying on an external (formal) system to provide the final rendering. There are some differences in the approaches that require some further consideration. 

\begin{description}
  \item[Context] In \tool{Isar}, the informal fragments are part of a formal document, which gives the context in which to evaluate the formal content. In our use case, there is no formal context: the informal and the formal documents are strictly separated, so the formal text has to exist already, and is only referred to from a natural-language document. 

  We could provide an extension that allows the author to specify the formal context in which formal text is evaluated. This would allow her to write new examples based on an existing formalization, or combine literate and non-literate approaches. This is an appealing idea, but beyond the scope of this paper.

  \item[Feedback] In our use case, the natural language text only refers to the formal text, and does not feed back any formal content into the formal document. In \tool{Isar}, it is possible to prove new lemmas in an antiquotation, but Wenzel notes in his thesis~\cite[page 65]{Wenzel-2002} that antiquotations printing well-typed terms, propositions and theorems are the most important ones in practice.
\end{description}

With these considerations in mind, we adopt the following syntax, based on the antiquotations: {\st @\{ type reference [options] \}}. The main element is {\st reference}, which is either a path in the Wiki or an external URL, pointing to a formal entity of the given {\st type}. We will discuss possible types in the next section. 

The {\st options} element instructs the renderer of the Wiki about how to render the included entity. Compared to \tool{Isar}, it has swapped positions with {\st reference} because it  provides rendering settings, and no instructions to a formal tool. This means that they are processed last, after the reference has been processed to an object. Possible uses include flagging whether or not to include the proof of a theorem, or the level of detail that should be shown when including a snippet.

The {\st reference} points at an annotated object, by giving the location of the document it occurs in and the name given in the annotation for that object. The {\st type} corresponds to the type in the annotation of the object: it serves as a disambiguation mechanism, but can be enforced in a more strict manner. If the system cannot find a reference of the given type, it should fail in a user friendly way: in our implementation, we inline {\st reference} in the output, marked up to show it is not found. Inspired by MediaWiki, we color it red, and put a question mark after it. An addition to this would be to make this rendering a link, through which the author can write the formal reference, or search for similar objects. 

The antiquotation for the Coq code in Figure~\ref{fig:result}, is \newline {\st @\{oo:Definition CoqBinomialCoefficient\#C\}}. It points to the {\st Definition} {\st C}, found in the location (a Wiki page) {\st CoqBinomialCoefficient}. This reference gets resolved into the HTML shown in the screenshot.

\section{Annotation of Types and Content}
\label{sec:types}

For transforming antiquotations to HTML, we could implement ad hoc reference resolution mechanisms specific to particular formal systems.  Then, any new formal system would require building another specific dereferencing implementation from scratch.  We present a more scalable approach with lower requirements for formal systems.
We enrich the HTML export of the formal texts with annotations, which clearly mark the elements that authors can refer to.  
The Wiki can resolve them in a uniform way: when an author writes an antiquotation, the system can dereference it to the annotated HTML, without further requirements on the structure of the underlying formal texts.

This section introduces the two main kinds of annotations that are relevant here; the next section explains how to put them into formal texts.  We are interested in annotating an item of formalized mathematics with its mathematical \emph{type} (such as definition, theorem, proof), and annotating it by pointing to related \emph{content} (such as pointing from a formalized proof to the Wikipedia article that gives an informal account of the same proof).  Type annotation requires a suitable annotation \emph{vocabulary}, whereas we had to identify suitable \emph{datasets} as targets for content annotation.

\subsection{The Type Vocabulary of the OMDoc Ontology}
\label{sec:type-vocab-omdoc}

The OMDoc ontology provides a wide supply of types of mathematical knowledge items, as well as types of \emph{relations} between them, e.g.\ that a proof proves a theorem~\cite{OMDocDocOnto:web,Lange:OntoLangMathSemWeb}.  It is a reimplementation of the conceptual model of the OMDoc XML markup language~\cite{Kohlhase:omdoc1.2} for the purpose of providing semantic Web applications with a vocabulary of structures of mathematical knowledge.\footnote{We use the terms ``ontology'' and ``vocabulary'' synonymously.}
It is thus one possible vocabulary (see \cite{Lange:OntoLangMathSemWeb} for others) applicable to the lightweight annotation of mathematical resources on the Web desired here, without the need to translate them from their original representation to OMDoc XML.

The OMDoc language has originally been designed for exchanging formalizations across systems for, e.g., structured specification, automated verification, and interactive theorem proving~\cite{Kohlhase:omdoc1.2}.  OMDoc  covers a large subset of the concepts of common languages for formalized mathematics, such as \tool{Mizar} or \tool{Coq}; in fact, partial translations of the latter languages to OMDoc have been implemented (see, e.g.,~\cite{BanKoh:mmlof07}).

The OMDoc \emph{ontology} covers most of the concepts that the OMDoc language provides for mathematical statements, structured proofs, and theories.  Item types include \textit{Theory}, \textit{Symbol} [Declaration], \textit{Definition}, \textit{Assertion} (having subtypes such as \textit{Theorem} or \textit{Lemma}), and \textit{Proof}; types of relations between such items include \textit{Theory--homeTheoryOf--<any type of statement>}, \textit{Symbol--hasDefinition--Definition}, and \textit{Proof--proves--Theorem}
.  The ontology leaves the representation of document structures without a mathematical semantics, such as sections within a theory that have not explicitly been formalized as subtheories, to dedicated document ontologies (cf.~\cite{Lange:OntoLangMathSemWeb}).

\subsection{Datasets for Content Annotation}
\label{sec:datas-cont-annot}

Our main use case for content annotation is annotating formalizations with related informal representations, but added-value services may still benefit from the latter having a \emph{partial} formal semantics.  Consider linking a formalized proof to a Wikipedia article that explains a sketch and the historical or application context of the proof.\footnote{The Wikipedia category ``Article proofs'' lists such articles; see \url{http://en.wikipedia.org/wiki/Category:Article_proofs}.}  The information in the Wikipedia article (such as the year in which the proof was published) is not immediately comprehensible to Web services or search engines.  For this purpose, DBpedia\fnurl{http://dbpedia.org} makes the contents of Wikipedia available as a linked open dataset.\footnote{A collection of RDF descriptions accessible by dereferencing their identifiers~\cite{HB:LinkedData11}}

Further suitable targets for content annotation of mathematical formalizations – albeit not yet available as machine-comprehensible linked open data – include the PlanetMath encyclopedia
, the similar ProofWiki
, and Wolfram's MathWorld.
\footnote{See \url{http://www.planetmath.org}, \url{http://www.proofwiki.org}, and \url{http://mathworld.wolfram.com}, respectively.}

In the interest of machine-comprehensibility, the links from the annotated sources to the target dataset should be \emph{typed}.  The two most widely used link types, which are also widely supported by linked data clients, are \textit{rdfs:seeAlso} (a generic catch-all, which linked data clients usually follow to gather more information) and \textit{owl:sameAs} (asserting that all properties asserted about the source also hold for the target, and vice versa).  The OMDoc ontology furthermore provides the link type \textit{formalizes} for linking from a formalized knowledge item to an informal item that verbalizes the former, and the inverse type \textit{verbalizes}.  

\section{Annotating Formal Texts}
\label{sec:annotating}

Now that we have established \emph{what} to annotate formal texts with, we need to look at the \emph{how}. 
Considering that the formal documents are stored on the Web, we assume that each document has an HTML representation. Indeed, the systems we support in our prototype each have some way of generating appropriate type annotations.

Text parts are annotated by enclosing them into HTML elements that carry the annotations as RDFa annotations.  RDFa is a set of attributes that allows for embedding RDF graphs into XML or HTML~\cite{AdidaEtAl08:RDFa}.  For identifying the annotated resources by URIs, as required by RDF, we reuse the identifiers of the original formalization.

\paragraph{Desired Results.}
Regardless of the exact details of the formal systems involved, and their output, the annotation process generally yields HTML+RDFa, which uses the OMDoc ontology (cf.\ Section~\ref{sec:type-vocab-omdoc}) as a vocabulary.  For example, if the formal document contains an HTML rendition of the Binomial Theorem, we expect the following result (where the prefix \textit{oo:} has been bound to the URI of the OMDoc ontology\footnote{\url{http://omdoc.org/ontology\#}}):

\begin{lstlisting}[language=XHTML+RDFa,basicstyle=\ttfamily\scriptsize]
<span typeof="oo:Theorem" about="#BinomialTheorem">...</span>
<span typeof="oo:Proof"><span rel="oo:proves" resource="#BinomialTheorem"/>
...</span>
\end{lstlisting}
The ``\ldots'' in this listing represent the original HTML rendition of the formal text, possibly including the information that was used to infer the annotations now captured by the RDFa attributes.  {\st @about} assigns a URI to the annotated resource; here, we use fragment identifiers within the HTML document. 

In this example, we wrap the existing HTML in {\st span} elements, because in most cases, this preserves the original rendering of the source text.  In particular, empty {\st span}s, as typically used when there is no other HTML element around that could reasonably carry some RDFa annotation, are invisible in the browser.  If the HTML of the source text contains {\st div} elements, it becomes necessary to wrap the fragment in a {\st div} instead of a {\st span}.

\paragraph{\tool{Mizar} texts.} 
Mizar processing consists of several passes, similar in spirit to
those used in compilation of languages like
Pascal and C. The communication between the main three passes
(parsing, semantic analysis, and proof checking) is likewise
file-based.  Since 2004, Mizar has been using XML as its native format
for storing the result of the semantic analysis~\cite{Urban05-mkm}. This
XML form has been since used for producing disambiguated
(linked) HTML presentation of Mizar texts, translating Mizar texts to
ATP formats, and as an input for a number of other systems. The use of
the XML as a native format guarantees that it remains up-to-date and
usable for such external uses, which has been an issue with a number
of ad-hoc ITP translations created for external use.

This encoding has
been gradually enriched  to contain important presentational
information (e.g., the original names of variables, the original
syntax of formulas before normalization, etc.), and also to contain additional
information that is useful for understanding of the Mizar texts, and
ATP and Wiki functions~\cite{UrbanS10,Urban+-2010} over them
(e.g.,
showing the thesis computed by the system after each natural deduction step,
linking to ATP calls/explanations, and section editing in a Wiki).

We implemented the RDF annotation of Mizar articles as a part of the XSL transformation
that creates HTML from the Mizar semantic XML format. While the OMDoc
ontology defines vocabulary that seems suitable also for many Mizar
internal proof steps, the current Mizar implementation only annotates
the main top-level Mizar items, together with the top-level
proofs. Even with this limitation this has already resulted in about
160000 annotations exported from the whole MML,\footnote{MML version
  4.178.1142 was used, see
  \url{http://mizar.cs.ualberta.ca/~mptp/7.12.02_4.178.1142/html/}}
which is more than enough for testing the Agora system. The
existing Mizar HTML namespace was re-used for the names of
the exported items, such that, for example, the Brouwer Fixed Point
Theorem:\footnote{\url{http://mizar.cs.ualberta.ca/~mptp/7.12.02_4.178.1142/html/brouwer.html\#T14}}
\begin{lstlisting}[language=Mizar,basicstyle=\ttfamily\scriptsize]
:: $N Brouwer Fixed Point Theorem
theorem Th14:
  for r being non negative (real number), o being Point of TOP-REAL 2, 
      f being continuous Function of Tdisk(o,r), Tdisk(o,r) 
  holds f has_a_fixpoint
proof ...
\end{lstlisting}

gets annotated 
as\footnote{{\st T14} is a \emph{unique internal} Mizar identifier denoting the theorem. {\st Th14} is a (possibly non-unique) \emph{user-level} identifier (e.g., {\st Brouwer} or {\st SK300} would result in {\st T14} too).}

\begin{lstlisting}[language=XHTML+RDFa,basicstyle=\ttfamily\scriptsize]
<div about="#T14" typeof="oo:Theorem">
  <span rel="owl:sameAs" 
        resource="http://dbpedia.org/resource/Brouwer_Fixed_Point_Theorem"/> ...
  <div about="#PF23" typeof="oo:Proof"><span rel="oo:proves" resource="#T14"/> ... </div>
</div>
\end{lstlisting}

Apart from the appropriate annotations of the theorem and
its proof, an additional \emph{owl:sameAs} link is
produced to the DBpedia (Wikipedia) ``Brouwer\_\allowbreak Fixed\_\allowbreak Point\_\allowbreak Theorem''
resource. Such links are produced for all Mizar theorems and concepts
for which the author defined a long (typically well-known) name using
the Mizar {\st ::\$N} pragma. Such pragmas provide a way for the
users to link the formalizations to Wikipedia (DBpedia, ProofWiki,
PlanetMath, etc.), and the links allow the data consumers (like Agora) to
automatically mesh together different (Mizar, Coq, etc.)
formalizations using DBpedia as the common namespace.

\paragraph{\Coq\ texts.} \Coq\ has access to type information when verifying a document. This information is written into a \emph{globalization} file, which lists types and cross-references on a line/character-offset basis. \Coq's HTML renderer, \tool{Coqdoc}, processes this information to generate hyperlinks between pages, and style parts of the document according to the given types. \tool{Coqdoc} is implemented as a single-pass scanner and lexer, which reads a \Coq\ proof script and outputs HTML (or \LaTeX) as part of the lexing process.

The resulting HTML page contains the information we defined in Section~\ref{sec:types}, but serves this in an unstructured way: individual elements of the text get wrapped in {\st span} elements corresponding to their syntactical class, and there is no further grouping of this sequence of {\st span}s in a more logical entity (e.g.\ a \lstinline[language=HTML,columns=fixed,breaklines=true,breakatwhitespace=true]{<div id="poly_id" class="lemma">...</div>}), which would be addressable from our syntax. In particular, it puts the name anchor around a theorem's label, instead of around the entire group.

For example, the following \Coq\ code:

\begin{lstlisting}[language=Coq,basicstyle=\ttfamily\scriptsize]
  Lemma poly_id: forall a, a -> a. 
\end{lstlisting}

gets translated into the following HTML fragment (truncated for brevity and whitespace added for legibility):

\begin{lstlisting}[language=HTML,deletekeywords={type},basicstyle=\ttfamily\scriptsize]
<span class="id" type="keyword">Lemma</span> 
<a name="poly_id"><span class="id" type="lemma">poly_id</span>
</a>...
\end{lstlisting}

\lstset{language = Coq}
Aside from the fact that the HTML is not valid ({\st span} elements do not allow {\st @type} attributes), it has the main ingredients we are interested in extracting for annotation (type and name), but no indication that the keyword \lstinline{Lemma}, the identifier \lstinline{poly_id} following it, and the statement \lstinline{forall a, a->a.} are related. The problem worsens for proofs: blocks of commands are not indicated as a proof, and there is no explicit relation between a statement and its proof, except for the fact that a proof always directly follows a statement.
This makes the \tool{Coqdoc}-generated HTML not directly suitable for our purpose; we need three steps of post-processing:

\begin{enumerate}
\item \textbf{Group objects:} The first step we take is grouping the `forest' of markup elements that constitutes a command for \Coq\ in a single element. This means parsing the text within the markup, and gathering the elements containing a full command in a new element.
\item \textbf{Export type information:} We then export the type information from \Coq\ to the new element. We extract this information from the {\st @type}, derive the corresponding OMDoc type from it, and put that into an RDFa {\st @typeof} attribute.
\item \textbf{Explicit subject identification:} The final step is extracting {\st @name} and putting it into {\st @about}, thus reusing it as a subject URI.
\end{enumerate}
After post-processing, we obtain the desired annotated tree, containing the HTML generated by \tool{Coqdoc}.

The approach introduced here does not yet allow us to indicate the proof blocks, for which we do need to modify \tool{Coqdoc}. The adaption is fairly straightforward: each time the tool notices a keyword starting a proof, it outputs the start of a new span, \lstinline[language=XHTML+RDFa]{<span typeof="oo:Proof">}. 
Similarly, the adapted tool outputs \lstinline[language=XHTML+RDFa]{</span>} when encountering a keyword signaling the end of a proof.

\paragraph{\tool{Isar} texts.} For \tool{Isar} texts, the annotation process is still in development. We make use of the \tool{Isabelle/Scala}~\cite{Wenzel-2011} library to generate HTML pages based on the proof structure, already containing the annotation of a page. Because the process has access to the full proof structure, it is easy to generate annotations: the main obstacle is that the information about the identifiers at this level does not distinguish between declaration and use, so it is difficult to know what items to annotate with an {\st @about}.

\section{System: Agora}
\label{sec:system}

\lstset{language=XHTML+RDFa}

We have implemented the mechanisms described in this paper as part of the Agora prototype.\footnote{\url{http://mws.cs.ru.nl/agora}} A current snapshot of the source can be found in our code repository.\footnote{\url{https://bitbucket.org/Carst/agora}}
Agora provides the following functionality, grouped by the tasks in the main use case. Writing and rendering the narrative is illustrated by the Agora page about the binomial coefficient.\footnote{\url{http://mws.cs.ru.nl/agora/cicm_sandbox/BinomialCoefficient}}

\paragraph{Formalization.} 
Agora allows the author to write her own formalizations grouped in \emph{projects}, which resemble repositories of formal and informal documents. Agora has some support for verifying \Coq\ formalizations, with a rudimentary editor for changing the files. Alternatively, it allows an author to synchronize her working directory with the system (currently, write access to the server is required for this). Proof scripts from this directory are picked up, and provided as documents. 
Agora also scrapes \tool{Mizar}'s MML for HTML pages representing theories, and includes them in a separate project.
Regardless of origin and editing methods, the proof scripts are rendered as HTML, and annotated using the vocabulary specified in Section~\ref{sec:types}.

\paragraph{Narratives.} To allow the author to write natural language narratives, we provide the Creole Wiki syntax~\cite{Sauer+-2007}, which allows an author to use a lightweight markup syntax. Next to this markup and the antiquotation described next, the author can write formulae in \LaTeX\ syntax, supported by the MathJax\fnurl{http://www.mathjax.org} library.

\paragraph{Antiquotation.} The author can include formal content from any annotated page by using the antiquotation syntax, just giving page names to refer to pages within Agora, or referring to other projects or URLs by writing a reference of the form: {\st @\{type location\#name\}}. For example, the formalization of the binomial coefficient in \Coq\ is included in the Wiki, so it can be referred to by {\st @\{oo:Definition CoqBinomialCoefficient\#C\}}. On the other hand, the \tool{Mizar} definition is given at an external Web page. Because the URL is rather long, the antiquotation is {\st @\{oo:Definition mml:binom.html\#D22\}}. In this reference, {\st mml} is a prefix, defined using the Agora-specific command {\st @\{prefix mml=\url{http://mizar.cs.ualberta.ca/~mptp/7.12.02_4.178.1142/html}\}}. We do not consider prefixes a part of the ``core'' syntax, as another implementation could restrict the system to only work within a single Wiki, causing the links to be (reasonably) short.
\\\\
\noindent Rendering a page written in this way, Agora transforms the Wiki syntax into HTML using a modified Creole parser. The modification takes the antiquotations and produces a placeholder {\st div} containing the type, reference and repository of the antiquotation as attributes. When the page is loaded, the placeholders are replaced asynchronously, by the referred-to entities. This step is necessary to prevent very long loading times on pages referring to many external pages. When the content is included, it is pre-processed to rewrite relative links to become absolute (with the source page used as the base URL), a matter of a simple library call.

An alternative to asynchronously fetching the referenced elements would be to cache them when the page is written. This approach could be combined with the asynchronous approach implemented, and would allow authors to refer to content that would, inevitably, disappear. We currently have not implemented it, because it requires some consideration in the scope of Agora's storage model.

\paragraph{Appearance of Included Content.} 
The appearance of included snippets depends on several things:

\begin{itemize}
  \item Cascading Style Sheets (CSS) are used to apply styling to objects that have certain attributes. When including the snippets, they can either be styled using the information in the source document (because the syntax is marked up according to rules for a specific system), or the styling can be specified in the including document (to make the rendered document look more uniform). 

In the implementation, we statically include the CSS files from the source text. This is manageable due to the small number of included systems, but requires further consideration.

  \item The system could use the included snippets to present the data using an alternative notation than the plain text that is typical for interactive theorem provers. This approach would require some system-dependent analysis of the included snippets, maybe going further than just HTML inclusion. The gathered data could then be used to render the included format in a new way, either specified by the author of the source text, or the author of the including text. This approach of ``re-rendering'' structured data was taken as part of the MoWGLI project~\cite{Asperti+-2005}, where the author of an informal text writes it as a \emph{view} on a formal structure, including transformations from the formal text to (mathematical) notation. 

  Because our approach intends to include a wide variety of HTML-based documents, we do not consider this notational transformation viable in general: it requires specific semantic information provided by the interactive theorem prover, which is not preserved in the annotation process described in Section~\ref{sec:annotating}, possibly not even exposed by the prover. However, where we have this information available, it would be good to use it to make a better looking rendered result.
\end{itemize}

\section{Conclusions and Future Work}
\label{sec:future}
This paper describes a mechanism for documenting formal proofs in an informal narrative. The narrative includes \emph{pointers} to objects found in libraries of formalized mathematics, which have been \emph{annotated} with appropriate types and names. The mechanism has been implemented as part of the Agora prototype.  Our approach is Web-scalable in that the Agora system is independent from a particular formalized library: It may be installed in a different place, it references formal texts by URL, and it does not make any assumptions about the system underlying the library, except requiring an HTML+RDFa export.  As future work, we see several opportunities for making the mechanisms more user friendly:

\paragraph{Include more systems.} By including more systems, we increase the number of objects an author can refer to when writing a Wiki page. Because we made our annotation framework generic, this should not be a very difficult task, and single documents could be annotated by authors on the fly, if necessary.
\paragraph{Provide other methods for reference.} At the moment, the {\st reference} part of our antiquotations is straightforward: the author should give a page and the identifier of the object in this page. It would be interesting to allow the author to use an (existing) query language to describe what item she is looking for, and use this to find objects in the annotated documents.
\paragraph{Improve editing facilities.} Agora currently has a simple text box for editing its informal documents. We could provide some feedback to the author by showing a preview of the marked up text, including resolved antiquotations. More elaborately, we could provide an `auto-completion' option, which shows the possible objects an author can refer to, limited by {\st type} and the partial path: if the author writes {\st @\{oo:Theorem Foo\#A\}}, the system provides an auto-completion box showing all the theorems in the ``Foo'' namespace, starting with ``A''.  This lookup could be realized in a generic way, abstracting from the different formalizations, by harvesting the RDFa annotations into an RDF database (``triple store'') and implementing a query in SPARQL.
\paragraph{Consistency.} The current design of the mechanisms already provides a better robustness than just including objects by giving a location, but can still be improved to deal with objects changing names. A solution would be to give objects an unchanging identifier and a human-readable name, and storing the antiquotation as a reference to this identifier. When an author edits a document containing an antiquotation, the name is looked up, and returned in the editable text. 
\paragraph{}
\noindent Despite these shortcomings, we believe we have made significant steps towards a system in which authors can document formal mathematics by pointing and writing, without having to commit prematurely to a specific workflow, such as literate proving, or even a tool chain, because representations of (formalized) mathematics can be annotated after they have been generated.

\bibliographystyle{plain}
\bibliography{kwarc,bibliography}
\end{document}